\documentstyle{article}

\normalsize

\begin{document}

\bibliographystyle{unsrt}
\footskip 1.0cm
\thispagestyle{empty}
\setcounter{page}{0}
\begin{flushright}
IC/94/65\\

Napoli: DSF-T-5/94, INFN-NA-IV-5/94\\
March 1994\\
\end{flushright}
\vspace{10mm}

\centerline {\LARGE Z FLUX-LINE LATTICES AND SELF-DUAL EQUATIONS}
\vspace{5mm}
\centerline {\LARGE IN THE STANDARD MODEL}
\vspace*{15mm}
\centerline {\large G. Bimonte}
\vspace{5mm}
\centerline{ \small and }
\vspace*{5mm}
\centerline{ \large G. Lozano \footnote{
E-mail addresses: Bimonte@ictp.trieste.it~~,~~Lozano@ictp.trieste.it}}
\vspace*{5mm}
\centerline {\it International Centre for Theoretical Physics, P.O.BOX 586}
\centerline {\it I-34100 Trieste, ITALY}
\vspace*{25mm}
\normalsize
\centerline {\bf Abstract}
\vspace*{5mm}
{\large
We derive gauge covariant self-dual equations for the
$SU(2) \times U(1)_Y$ theory of electro-weak interactions and show that
they admit solutions describing a periodic lattice of Z-strings.} \newpage

\baselineskip=24pt
\setcounter{page}{1}

Solutions to the classical equations of motion of the Standard Model have
recently attracted a
considerable attention after the observation made by Vachaspati
\cite{vacha} that the embedding \cite{nambu}
of the Nielsen-Olesen~\cite{niel} string into the Standard Model
can be
classically stable in a certain range of parameters. Further studies
\cite{vacha2} have revealed that the region of classical stability does
not overlap with the one compatible with experimental data.
It has been suggested \cite{perki} that the cause of
this instability might be traced back to the phenomenon of W-condensation
\cite{ambjo}
but the search of stable W-dressed strings has given negative results
\cite{achuc}.

In their original investigations on electroweak magnetism, Ambj$\o$rn and
Olesen \cite{ambjo} have solved the electroweak equations showing the
existence
of classically stable configurations representing a condensate of
$W$ bosons. Classical stability is ensured since these configurations
satisfy a set of first order differential equations (self-dual equations)
which arise from the requirement that they saturate a Bogomol'nyi like
bound \cite{bogo} for the energy.
The
solutions found in Ref. \cite{ambjo} carry no $Z$ flux. The reason for
this is that the self-dual equations were derived in the unitary gauge
which excludes from the
beginning the existence of $Z$-strings.

In this work we present a
 {\it gauge covariant} expression of the self-dual equations and show
that
there exist boundary conditions which lead to configurations
carrying $Z$ flux.

We will
consider static configurations which are axially symmetric along the
$z$-axis and such that physical observables are periodic in the $x-y$
plane. The energy functional for a cell $\cal C$ of periodicity with
area $A$ is then
given by
\begin{equation}
E=\int_{\cal C} d^2x \left[\frac{1}{4} W^a_{ij}W^a_{ ij}+
\frac{1}{4} B_{ij}B_{ij}
+ (D_{i}\Phi)^{\dagger}D_i\Phi
 + \lambda (\Phi^{\dagger} \Phi - \phi_0^2)^2\right]~~, \label{1}
\end{equation}
where $W^a_{ij}$ and $B_{ij}$ are the field
strengths associated with the gauge group
$SU(2) \times U(1)_Y$
$$
W^a_{ij}=
\partial_{i}W^a_{j}-\partial_{j}W^a_{i}
+g \epsilon_{abc}W^b_{i}W^c_{j}~~,
$$
\begin{equation}
B_{ij}=\partial_{i}B_{j}-\partial_{j}B_{i}~~,
\end{equation}
while the covariant derivative
for the Higgs-doublet $\Phi$ is defined as
\begin{equation}
D_{i}\Phi=\left( \partial_{i} - \frac{i}{2}g \tau^a W^a_{i}-
\frac{i}{2}g^{\prime} B_{i}
\right) \Phi .     \label{2}
\end{equation}
In Eq.(3), $\tau^a$ are the Pauli matrices and $g$ and $g^{\prime}$ are
related to the Weinberg angle $\theta_W$ and to the electric charge $e$
by the relations
\begin{equation}
e=g \sin \theta _W~~~~~~~g^{\prime}=g \tan \theta_W.
\end{equation}
The electromagnetic field $A_{i}$, the neutral $Z_{i}$ field and the
charged $W_{i}$ field are defined as:
$$
A_{i}=\sin \theta_W W^3_{i}+\cos \theta_W B_{i}~~,
$$
$$
Z_{i}=\cos \theta_W W^3_{i}-\sin \theta_W B_{i}~~,
$$
\begin{equation}
W_{i}=\frac{1}{\sqrt 2}(W^1_{i}-iW^2_{i})~~.
\end{equation}
and the masses of the physical particles are given by:
\begin{equation}
m^2_W=\frac{1}{2}g^2 \phi_0^2~,~~~m^2_Z=\frac{g^2 \phi_0^2}{2 \cos^2
\theta_W}~, ~~~m^2_H=4 \lambda \phi^2_0~~.
\end{equation}

In order to find a bound for the energy, we will start by rewriting the
kinetic energy of the Higgs field in an appropriate way by extending to
the electroweak theory the identity used by Bogomol'nyi in the simpler
$U(1)$ Landau-Ginzburg theory \cite{bogo}; this identity reads:
\begin{eqnarray}
\left| D_{i}\Phi
\right|^2 & = & \frac{1}{2}\left| D_{i}\Phi -i \gamma
\epsilon_{ij}D_{j}\Phi\right|^2 + \nonumber\\
&  & -\frac{1}{4}\gamma g \Phi^{\dagger}
\tau^a \Phi \epsilon_{ij}W^a_{ij}-
\frac{1}{4}\gamma g^{\prime}
 \Phi^{\dagger}
 \Phi \epsilon_{ij}B_{ij}+ \gamma\epsilon_{ij} \partial_i J_j \label{3}
\end{eqnarray}
where
\begin{equation}
J_j=\frac{1}{2i}\left[ \Phi^{\dagger}D_j \Phi-
(D_j \Phi)^{\dagger}\Phi \right] \label{4}
\end{equation}
and $\gamma^2$=1 In what follows we will set $\gamma=-1$ (the opposite case
can be treated analogously).
After making use of Eq.(7), the energy can be written as:
\newpage
$$
E=E_1+ \frac{ g^{\prime} \phi_0^2}{4 \sin^2 \theta_W}\int_{\cal C}
d^2x \epsilon_{ij}B_{ij}- \frac{g^2 \phi_0^4}{8 \sin^2 \theta_W}\int_
{\cal C} d^2x +\left( \lambda-\frac{g^2}{8 \cos^2 \theta_W}\right)\int_
{\cal C}d^2x (\Phi^{\dagger}\Phi-\phi_0^2)^2+
$$
\begin{equation}
- \int_{\cal C}
d^2x \epsilon_{ij}\partial_i J_j~~, \label{5}
\end{equation}
where $E_1$ is the positive definite functional
$$
E_1 =  \int_{\cal C} d^2x  \left\{ \frac{1}{4}\left[W^a_{ij}+\epsilon_{ij}
 \frac{g}{2}\Phi^{\dagger}\tau^a \Phi \right]^2+\frac{1}{4}
\left[B_{ij}+\epsilon_{ij} \frac{g^{\prime}}{2}\left(\Phi^{\dagger}\Phi-
\frac{\phi_0^2}{\sin^2 \theta_W}\right)
\right]^2+ \right.
$$
\begin{equation}
 + \left. \frac{1}{2}\left| D_{i}\Phi
+i \epsilon_{ij}D_{j}\Phi\right|^2 \right\} ~~.\label{6}
\end{equation}
The last term in Eq.(9) can always be dropped after integration on a
cell, since it is the divergence of a gauge invariant quantity
(remember that observables are periodic). The second term in (9) is of
topological nature and is proportional to the flux $\Phi_B$ of the $U(1)_Y$
field $B_i$, while the third term is a geometric term proportional to the
area $A$ of the cell. We then see from (9) that for
\begin{equation}
\lambda \ge \frac{g^2}{8 \cos^2 \theta_W} \label{7}
\end{equation}
or equivalently
\begin{equation}
m_H \ge m_Z~~,
\end{equation}
the energy is bounded by
\begin{equation}
E_{\rm {bound}}=\frac{ g^{\prime}\phi_0^2}{2 \sin^2 \theta_W}\Phi_B
-\frac{g^2 \phi_0^4}{8 \sin^2 \theta_W}A=
\frac{m^2_W}{2 e^2}\left(2 g^{\prime} \Phi_B - m^2_W A \right) \label{8}.
\end{equation}
This is the Bogomol'nyi bound for the Standard Model, derived in a gauge
covariant way. The simpler $U(1)$ case is recovered by setting $g=0$ in
Eq.(13). A significant
difference between the abelian and the non-abelian case is the presence
in the latter of the term proportional to the area, which spoils the
topological nature of the bound.
Although (12) covers most of the experimental region ($m_H > 0.62 m_Z$),
let us point out that a slight modification of (7) as in Ref. \cite{cugli}
allows one to deduce an expression of the bound valid in the case
$m_H < m_Z$:
\begin{equation}
E_{\rm bound}=\frac{1}{2 e^2}
\left( \frac{m_W m_H}{m_Z}\right)^2
\left[2 g^{\prime} \Phi_B- \left( \frac{m_W m_H}{m_Z}\right)^2 A
\right]~~.
\end{equation}

Now, let us concentrate on configurations saturating the bound. This occurs
only if $m_H=m_Z$ and $E_1=0$, which in turn implies:
\begin{eqnarray}
W^a_{ij} & = &-\epsilon_{ij} \frac{g}{2}\Phi^{\dagger}\tau^a \Phi~~~,
\label{11}\\
B_{ij} & = & -\epsilon_{ij}\frac{g^{\prime}}{2}\left(\Phi^{\dagger}\Phi-
\frac{\phi_0^2}{\sin ^2 \theta_W}\right)  ~~~,
\label{12}\\
D_i \Phi & = & -i  \epsilon_{ij}D_j\Phi~~~.   \label{13}
\end{eqnarray}
This is the gauge covariant form of the self-dual (or Bogomol'nyi)
equations
previously found in the
unitary gauge by Ambj$\o$rn and Olesen \cite{ambjo}.

Of course, for $g=0$, Eqs.(15-17) become the well studied self-dual equations
of the $U(1)$ model \cite{bogo,jaffe,weinb}. For the purely non-abelian
case, which is obtained
by taking $g^{\prime}=0$, it is in some way surprising that the same
equations appear in a quite different context, that of a theory of
non-relativistic bosons interacting with a Chern-Simons field \cite{dunne}.

We will now look for solutions of (15-17), corresponding
to periodic observables in the plane. In
order to simplify the
discussion of boundary conditions it is convenient to point out that every
solution of
Eqs. (15-17) can be converted via a smooth
gauge transformation into
the {\it generalized} unitary gauge
\begin{equation}
\Phi= \left(\matrix{ 0 \cr \phi \cr}\right)~~~. \label{14}
\end{equation}
where $\phi$ is {\it complex}. This follows from the fact
that all Higgs doublets satisfying the Bogomol'nyi Eq.(17) can be
written locally as:
\begin{equation}
\Phi=(z-z_0)^n \Psi(x,y)~~, \label{15}
\end{equation}
where $z=x+iy$ and $\Psi(x,y)$ is a doublet such that:
$$
(\Psi^{\dagger}\Psi) (z_0)\neq 0.
$$
It then follows that $\Psi$ can be converted to the unitary gauge locally.
This transformation can always be extended globally to the plane
without obstructions. The most general boundary conditions which give
periodic observables
are then (for simplicity we consider a rectangular standard
cell of sides $L_X$ and $L_Y$):
\begin{eqnarray}
\Phi(x+L_X, y) & = & e^{i[\alpha_X(x,y) \tau_3+\beta_X(x,y)]}\Phi(x,y)
\nonumber\\ W_i(x+L_X,y) &= & e^{2i\alpha_X(x,y)}W_i(x,y)\nonumber \\
W^3_i(x+L_X,y) &= & W^3_i(x,y)+\frac{2}{g}\partial_i\alpha_X(x,y)\nonumber\\
B_i(x+L_X,y) & = & B_i(x,y)+\frac{2}{g^{\prime}}\partial_i\beta_X(x,y),
\label{16} \end{eqnarray}
with analogous boundary conditions in the $y$ direction, with functions
$\alpha_Y$ and $\beta_Y$. It
is easily verified that $\alpha_Y$ and $\beta_Y$ can be gauged away and
we can
then assume without any loss of generality that all fields are periodic
in the $y$ direction. Consistency of the boundary conditions (20) at the
corners of the cell implies the following constraints for $\alpha_X$ and
$\beta_X$: $$
\alpha_X(x,y)=\frac{\pi k_3}{L_Y}y + {\tilde \alpha}_X(x,y)
$$
\begin{equation}
\beta_X(x,y)=\frac{\pi k}{L_Y}y + {\tilde \beta}_X(x,y)~~\label{17},
\end{equation}
where $k$ and $k_3$ are integers such that $k-k_3$ is even. The
functions ${\tilde  \alpha}_X$ and ${\tilde \beta}_Y$ are periodic
in $y$ and can be gauged away. Summarizing, the most general boundary
conditions compatible with Eqs. (15-17) read:
\begin{eqnarray}
\phi(x+L_X,y)&= &e^{i\pi(k-k_3)\frac{y}{L_y}}\phi(x,y)~~,\nonumber\\
W_i(x+L_X,y)& = &e^{2i\pi k_3\frac{y}{L_y}}W_i(x,y)~~,\nonumber\\
W^3_i(x+L_X,y) &= &W^3_i(x,y)+\frac{2 \pi
k_3}{gL_Y}\delta_{i,2}~~,\nonumber\\
B_i(x+L_X,y) &= &B_i(x,y)+\frac{2 \pi k}{g^{\prime}L_Y}\delta_{i,2}~~.
\label{18}
\end{eqnarray}
Boundary conditions (22) reduce to those considered in
\cite{ambjo} only if $k=k_3$.
In this case, $\phi$ is real and periodic.
Solutions with different values of $k$ and $k_3$ are gauge inequivalent.

The integers $k$ and
$k_3$ are related to the fluxes of $f_{12}$ and $Z_{12}$:
\begin{equation}
\Phi_A \equiv \int_{\cal C} d^2x f_{12}=\frac{2 \pi k}{e}-\frac{2 \pi \sin^2
\theta_W}{e}(k-k_3)~,\label{19}
\end{equation}
\begin{equation}
\Phi_Z \equiv \int_{\cal C} d^2x Z_{12}=-\frac{\pi}{e}\sin 2 \theta_W
(k-k_3)~~.\label{20} \end{equation}
Using Eq.(18) in Eq. (17), we now obtain the following equation for the
field $W_i$:
\begin{equation}
(W_i+i \epsilon_{ij}W_j)\phi=0~~,\label{22}
\end{equation}
which implies that, outside the zeroes of $\phi$,
\begin{equation}
W_1=-i W_2 \equiv W~~.\label{23}
\end{equation}
Using this relation and Eq. (18) into eqs (15-17), we get:
\begin{equation}
{\tilde D}_iW+i\epsilon_{ij}{\tilde D}_j W=0~~,\label{24}
\end{equation}
\begin{equation}
D_i \hat \phi+i\epsilon_{ij} D_j \hat \phi=0~~,\label{25}
\end{equation}
\begin{equation}
f_{12}=\frac{m^2_W}{e} + 2e |W|^2~~,\label{26}
\end{equation}
\begin{equation}
Z_{12}=\frac{m^2_W \tan \theta_W}{e}({\hat \phi}^* \hat \phi
-1)+\frac{2e}{\tan \theta_W} |W|^2~~,\label{27}
\end{equation}
where $\hat \phi =\phi/\phi_0$ and
$$
{\tilde D}_i=\partial_i-i e A_i - i \frac{e}{\tan \theta_W}Z_i~~,
$$
$$
D_i= \partial_i+i\frac{e}{\sin 2 ~\theta_W}Z_i~~.
$$
Eq. (27-30) have solutions only when $k_3$ and $k-k_3$ are non-negative
(the opposite statement holds when $\gamma=1)$.
Notice that the energy of configurations satisfying the Bogomol'nyi
equations depends only on $k$,
\begin{equation}
E=\frac{\phi^2_0}{\sin^2 \theta_W} \pi k - \frac{g^2 \phi^4_0}
{8 \sin ^2 \theta_W} A~~.\label{21}
\end{equation}
We then see that different values of $k_3$ correspond to degenerate but not
gauge equivalent
solutions.\\
These equations are consistent only if the area $A$ of the standard cell
belongs to a range determined by the fluxes and the parameters of the
theory. By integrating Eq. (29), we get:
\begin{equation}
e \Phi_A=m^2_W A +2 e^2 \int_{\cal C} d^2x |W|^2 \ge m^2_W A~~,\label{28}
\end{equation}
or
\begin{equation}
m^2_W A \le 2 \pi k - 2 \pi \sin^2 \theta _W (k-k_3)~~.\label{29}
\end{equation}
The lower bound on the area can be obtained directly from Eq.(16):
\begin{equation}
g^{\prime} \Phi_B =-g^{\prime 2}\int_{\cal C} d^2x
\Phi^{\dagger}\Phi + \frac{g^{\prime 2}\phi_0^2}{2 \sin ^2 \theta_W}A
\le \frac{g^{\prime 2}\phi_0^2}{2 \sin ^2 \theta_W}A~~,\label{30}
\end{equation}
or
\begin{equation}
m^2_W A \ge 2 \pi k \cos^2 \theta_W~~.\label{31}
\end{equation}
These bounds coincide with those derived in \cite{ambjo} when $k=k_3$.
Notice nevertheless that while the upper bound still depends only on the flux
of $f_{12}$, the lower bound now also depends on the flux of $Z_{12}$.
It is worth to analize the case when the area $A$ is equal to
the upper bound (33). In this case, we
obtain a solution by setting
$$
W=0~~,\label {32}
$$
\begin{equation}
f_{12}=\frac{m^2_W}{e}\equiv H_{c_1}~~.
\end{equation}
The equations for $\hat \phi$ and $Z_i$ become:
\begin{equation}
D_i \hat \phi + i \epsilon_{ij} D_j \hat \phi=0~~,\label{33}
\end{equation}
\begin{equation}
Z_{12}= \frac{m^2_W \tan \theta_W}{e}({\hat \phi} ^* \hat
\phi -1)~~.\label{34} \end{equation}
Eqs. (37) and (38) are the self-dual equations of a $U(1)$ theory defined
on a torus. Although no explicit form of the solutions is available, their
existence as well as the properties of the associated moduli space have
been studied in detail \cite{shah}. Notice nevertheless the presence of a
constant magnetic field (36) which is the remnant of the non-abelian
structure of the equations. This shows itself also in the expression
of the energy of these configurations which differs from the abelian
formula by a term proportional to the area, which is precisely the energy
of the electromagnetic field,
\begin{equation}
E=\frac{e \phi_0^2}{\sin 2 \theta_W}|\Phi_Z|+\frac{1}{2}H_c^2 A~~.
\end{equation}

We have then derived a gauge covariant expression for the self-dual
equations of the Standard Model and showed, in a particular case, the
existence of classically stable configurations
carrying $Z$ flux. As in Ref. \cite{vacha}, these configurations
correspond to an embedding of $U(1)$ into $SU(2)\times U_Y(1)$, but in this
case classical stability holds for values of the parameters not ruled out
by experiments ($m_H=m_Z$). Finally, our configurations present a
superimposed constant magnetic field, a phenomenon that can be linked to
the non-abelian nature of the equations. Notice also in this respect that
our solutions correspond to a minimization of the energy on a
{\it compact} region.

The structure represented by Eqs. (36-39) is simple but is far from
being general. It would be interesting to explicitly show the existence
of solutions for $f_{12}>H_{c_1}$, in which case a $W$-condensate will be
present. In this regard, we believe that the methods used in Ref.
\cite{spruck} can be applied after suitable modifications.

We would like to thank Prof. Abdus Salam, the International Atomic Energy
Agency and UNESCO for hospitality at the International Centre
for Theoretical Physics.

\end{document}